\documentclass[twocolumn,trackchanges]{aastex62}

\graphicspath{{./}{figures/}}
\acceptjournal{ApJL}
\usepackage{ulem}

\shorttitle{High-redshift galaxies in $\Lambda$CDM}
\shortauthors{Haslbauer et al.}

\begin{document}

\title{Has JWST already falsified dark-matter-driven galaxy formation?}

\correspondingauthor{Moritz Haslbauer}
\email{mhaslbauer@astro.uni-bonn.de}

\author[0000-0002-5101-6366]{Moritz Haslbauer}
\affiliation{Helmholtz-Institut f\"ur Strahlen- und Kernphysik (HISKP), University of Bonn, Nussallee 14$-$16, D-53115 Bonn, Germany}
\affiliation{Max-Planck-Institut f\"ur Radioastronomie, Auf dem H\"ugel 69, D-53121 Bonn, Germany}

\author[0000-0002-7301-3377]{Pavel Kroupa}
\affiliation{Helmholtz-Institut f\"ur Strahlen- und Kernphysik (HISKP), University of Bonn, Nussallee 14$-$16, D-53115 Bonn, Germany}
\affiliation{Astronomical Institute, Faculty of Mathematics and Physics, Charles University, V Hole\v{s}ovi\v{c}k\'ach 2, CZ-180 00 Praha 8, Czech Republic}

\author[0000-0002-0322-9957]{Akram Hasani Zonoozi}
\affiliation{Department of Physics, Institute for Advanced Studies in Basic Sciences (IASBS), Zanjan 45137-66731, Iran}

\author[0000-0002-9058-9677]{Hosein Haghi}
\affiliation{Department of Physics, Institute for Advanced Studies in Basic Sciences (IASBS), Zanjan 45137-66731, Iran}



\begin{abstract}
The James Webb Space Telescope (JWST) discovered several luminous high-redshift galaxy candidates with stellar masses of $M_{*}\ga 10^{9} \, M_{\odot}$ at photometric redshifts $z_{\mathrm{phot}} \ga 10$ which allows to constrain galaxy and structure formation models. For example, Adams et al. identified the candidate ID~1514 with $\log_{10}(M_{*}/M_{\odot}) = {9.8}_{-0.2}^{+0.2}$ located at $z_{\mathrm{phot}} = 9.85_{-0.12}^{+0.18}$ and Naidu et al. found even more distant candidates labeled as GL-z11 and GL-z13 with $\log_{10}(M_{*}/M_{\odot}) = 9.4_{-0.3}^{+0.3}$ at $z_{\mathrm{phot}}=10.9_{-0.4}^{+0.5}$ and $\log_{10}(M_{*}/M_{\odot}) = 9.0_{-0.4}^{+0.3}$ at $z_{\mathrm{phot}} = 13.1_{-0.7}^{+0.8}$, respectively. Assessing the computations of the IllustrisTNG (TNG50-1 and TNG100-1) and EAGLE projects, we investigate if the stellar mass buildup as predicted by the $\Lambda$CDM paradigm is consistent with these observations assuming that the early JWST calibration is correct and that the candidates are indeed located at $z~\ga~10$. Galaxies formed in the $\Lambda$CDM paradigm are by more than an order of magnitude less massive in stars than the observed galaxy candidates implying that the stellar mass buildup is more efficient in the early universe than predicted by the $\Lambda$CDM models. This in turn would suggest that structure formation is more enhanced at $z~\ga~10$ than predicted by the $\Lambda$CDM framework. We show that different star formation histories could reduce the stellar masses of the galaxy candidates alleviating the tension. Finally, we calculate the galaxy-wide initial mass function (gwIMF) of the galaxy candidates assuming the integrated galaxy IMF theory. The gwIMF becomes top-heavy for metal-poor starforming galaxies decreasing therewith the stellar masses compared to an invariant canonical IMF.
\end{abstract}

 

\section{Introduction} \label{sec:intro}
The formation of the first galaxies in the observed universe is a key question in modern astrophysics and one of the most important science goals of the recently launched James Webb Space Telescope (JWST). The Near Infrared Camera instrument \citep[NIRCam;][]{Rieke_2005} of JWST observes the universe in the $\approx 0.6-5 \, \mu m$ regime. This allows the detection of objects at redshifts $z \ga 12$, thus, revealing the evolutionary stage of galaxies in the early Universe just $\approx 400$~Myr after the Big Bang. 
The most distant confirmed galaxy is GN-z11 with a stellar mass of $M_{*} \approx 10^{9}\,M_{\odot}$ at a spectroscopic redshift of $z_{\mathrm{spec}} = 11.09^{+0.08}_{-0.12}$ detected with the Hubble Space Telescope \citep[][]{Oesch_2016}. Recently, \citet{Naidu_2022} reported the discovery of two luminous galaxy candidates labeled as GLASS-z11 (hereafter GL-z11) and GLASS-z13 (GL-z13) with $\log_{10}(M_{*}/M_{\odot}) = 9.4_{-0.3}^{+0.3}$ and $9.0_{-0.4}^{+0.3}$ located at photometric redshifts of $z_{\mathrm{phot}}=10.9_{-0.4}^{+0.5}$ and $z_{\mathrm{phot}} = 13.1_{-0.7}^{+0.8}$, respectively (see their table~3). In a subsequent publication, \citet{Naidu_2022b_dust} presented the galaxy candidate CEERS-1749 with $\log_{10}(M_{*}/M_{\odot}) = 9.6_{-0.2}^{+0.2}$ most likely located at $z_{\mathrm{phot}} = 16.0_{-0.6}^{+0.6}$ but a secondary redshift solution of $z \approx 5$ cannot be excluded (see their figure~1 and table~3).

Further luminous high-redshift galaxy candidates have been discovered by JWST \citep[e.g.,][]{Adams_2022,Atek_2022,Furtak_2022,Harikane_2022,Jan_2022_JWST,Labbe_2022}. For example, \citet{Adams_2022} studied the properties of candidates over a redshift range of $9 < z_{\mathrm{phot}} < 12$ and the source with ID~1514 has $\log_{10}(M_{*}/M_{\odot}) = {9.8}_{-0.2}^{+0.2}$ at $z_{\mathrm{phot}} = 9.85_{-0.12}^{+0.18}$ (see their tables~3 and 5). \citet{Labbe_2022} identified candidates with $M_{*} > 10^{10} \, M_{\odot}$ over $7 < z_{\mathrm{phot}} < 11$ from which the two most massive are ID~14924 with $\log_{10}(M_{*}/M_{\odot}) = 10.93$ at $z_{\mathrm{phot}} = 9.92$ and ID~38094 with $\log_{10}(M_{*}/M_{\odot}) = 11.16$ at $z_{\mathrm{phot}} = 7.56$ (see their figure~3). The spectroscopical confirmation of these objects is still outstanding because the high photometric redshifts of these galaxy candidates can, e.g., artificially emerge due to dust attenuation \citep{Naidu_2022,Naidu_2022b_dust, Zavala_2022}. 

In this contribution, we aim to investigate if ID~1514, ID~14924, GL-z11, GL-z13, and CEERS-1749 are consistent with the hierarchical buildup of stellar mass as predicted by the $\Lambda$CDM paradigm \citep{Efstathiou_1990,Ostriker_1995} using the Illustris The Next Generation \citep[TNG;][]{Pillepich_2018_TNG,Nelson_2019,Pillepich_2019} and Evolution and Assembly of GaLaxies and their Environments \citep[EAGLE;][]{Crain_2015,Schaye_2015,McAlpine_2016} projects.

\section{Method} \label{sec:Method}
The IllustrisTNG \citep[][]{Pillepich_2018_TNG,Pillepich_2018,Pillepich_2019,Nelson_2018,Nelson_2019,Nelson_2019_TNG50,Springel_2018,Naiman_2018,Marinacci_2018} and EAGLE \citep[][]{Crain_2015,Schaye_2015,McAlpine_2016} projects consist of a suite of hydrodynamical cosmological simulation runs conducted in the $\Lambda$CDM framework. 

The IllustrisTNG project assumes a Planck-2015 \citep{Planck_2016_IllustrisTNG} cosmology with the cosmological parameters being $H_{0} = 67.74\,\rm{km\,s^{-1}\,Mpc^{-1}}$, $\Omega_{\mathrm{b,0}} = 0.0486$, $\Omega_{\mathrm{m},0} = 0.3089$, $\Omega_{{\Lambda},0} = 0.6911$, $\sigma_{8} = 0.8159$, and $n_{\mathrm{s}} = 0.9667$. The simulation runs are based on the moving-mesh code AREPO \citep{Springel_2010} and self-consistently evolve the gas cells, stellar, black hole, and dark matter particles from redshift $z = 127$ up to present time. The Subfind Subhalos\footnote{\url{https://www.tng-project.org/data/docs/specifications/\#sec2b} [23.07.2022]} and Group\footnote{\url{https://www.tng-project.org/data/docs/specifications/\#sec2a} [12.09.2022]} catalogs are available for 100 different time steps (snapshots) in the redshift range of $0 \leq z \leq 20.05$. Here, we analyze the snapshot at redshift $z = 14.99$ (corresponding to an age of the universe of $t=0.271$~Gyr and a snapshot number of snapnum = 1), at $z = 11.98$ ($t=0.370$~Gyr, snapnum = 2), at $z = 10.98$ ($t = 0.418$~Gyr, snapnum = 3), and at $z = 10.00$ ($t = 0.475$~Gyr, snapnum = 4) of the high-resolution realization TNG50-1 and TNG100-1. The former has a box side of $35 \, h^{-1} = 51.7$ comoving Mpc (cMpc), respectively, where $h$ is the present Hubble constant $H_{0}$ in units of $100 \, \rm{km\,s^{-1}\,Mpc^{-1}}$, a baryonic element mass of $m_{\mathrm{b}} = 8.5 \times 10^{4} \, M_{\odot}$ and a dark matter particle mass of $m_{\mathrm{dm}} = 4.5 \times 10^{5} \, M_{\odot}$. TNG100-1 has, with $75 \, h^{-1} = 110.7$ cMpc, a larger box size and with $m_{\mathrm{b}} = 1.4 \times 10^{6} \, M_{\odot}$ and $m_{\mathrm{dm}} = 7.5 \times 10^{6} \, M_{\odot}$ a lower resolution than TNG50-1 \citep[see, e.g., also table 1 of][]{Nelson_2019}.

The EAGLE project is consistent with the Planck-2013 \citep{Planck_2014_EAGLE} cosmology being $H_{0} = 67.77\,\rm{km\,s^{-1}\,Mpc^{-1}}$, $\Omega_{\mathrm{b},0} = 0.04825$, $\Omega_{\mathrm{m},0} = 0.307$, $\Omega_{\Lambda,0} = 0.693$, $\sigma_{8} = 0.8288$, and $n_{\mathrm{s}} = 0.9611$ \citep[see also table~1 of][]{Schaye_2015}. Its simulations run with a modification of the GADGET-3 smoothed particle hydrodynamics code \citep[e.g.,][]{Springel_2005} starting also at $z = 127$ and self-consistently evolving the baryonic and dark matter particles up to the present day. The publicly available subhalo catalogs \citep[table~B.1 of][]{McAlpine_2016} are recorded for 29 snapshots in the redshift range of $0 \leq z \leq 20.00$ \citep[see table C.1 of][]{McAlpine_2016}. 

We use the two high-resolution realization runs RefL0025N0752 and RecalL0025N0752, and the two lower resolution runs RefL0050N0752 and RefL0100N1504 at $z = 15.13$ (snapnum = 1) and $z = 9.99$ (snapnum = 2). The two high-resolution runs have a box size of $25\,\rm{cMpc}$ with an initial baryonic particle mass of $m_{\mathrm{b}} = 2.26 \times 10^{5} \, M_{\odot}$, and a dark matter particle mass of $m_{\mathrm{dm}} =1.21 \times 10^{6} \, M_{\odot}$.
RefL0050N0752 and RefL0100N1504 have a size of $50$ cMpc and $100$ cMpc, respectively, and both have an initial baryonic particle mass of $m_{\mathrm{b}} = 1.81 \times 10^{6} M_{\odot}$ and a dark matter particle mass of $m_{\mathrm{dm}} = 9.70 \times 10^{6} M_{\odot}$ \citep[table~2 of][]{Schaye_2015}.
\newpage 
\section{Results} \label{sec:Results}
\begin{figure*}
    \includegraphics[width=\linewidth]{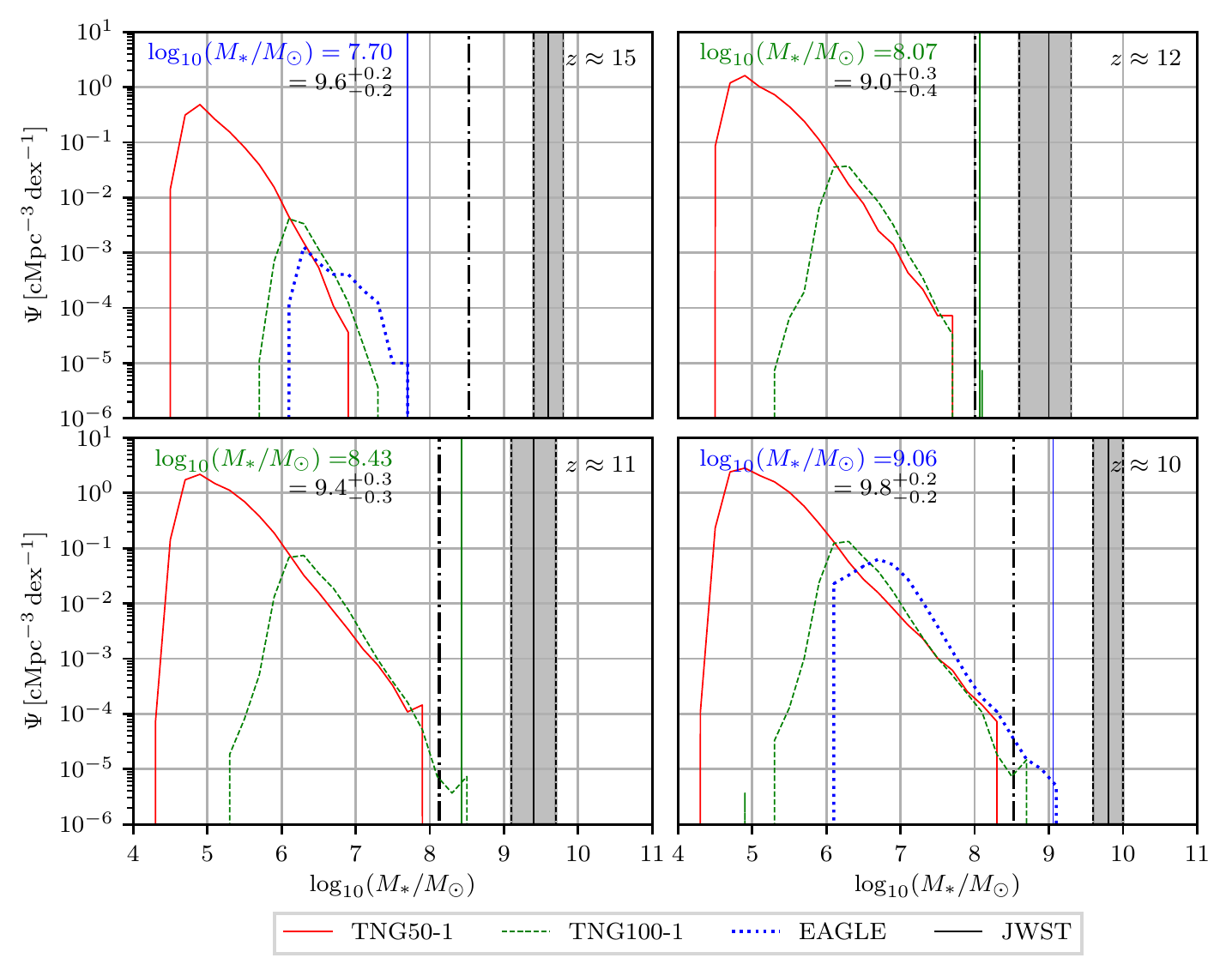}
    \caption{The GSMF at redshifts $z \approx 15$ (top left), $12$ (top right), $11$ (bottom left), $10$ (bottom right) in the TNG50-1 (solid red), TNG100-1 (dashed green), and RefL0100N1504 (dotted blue) simulation. The colored vertical solid line marks the most massive subhalo in terms of the stellar mass in the simulations. The vertical black solid lines refer to the reported galaxy candidates CEERS-1749 (top left), GL-z13 (top right), GL-z11 (bottom left), and ID~1514 (bottom right), where the black dashed lines correspond to the measurement uncertainties. The vertical dashed-dotted lines mark the lowest possible value as inferred for different SFHs in Section~\ref{subsec:minimum stellar mass for SFHs}. The histograms are normalized by their bin width of $\Delta \log_{10}( M_{*}/M_{\odot}) = 0.2$ and volume of the simulation box.}
    \label{figure_stellar_mass_distribution_LCDM}
\end{figure*}

The galaxy stellar mass function (GSMF) at redshifts $z = 14.99, 11.98$, $10.98$, and $10.00$ in the TNG runs and at $z = 15.13$ and $9.99$ in the EAGLE runs are presented in Figure~\ref{figure_stellar_mass_distribution_LCDM}.
The global peak of the distribution depends on the resolution and/or box size of the simulation runs such that the formation of low massive galaxies depends on the particle resolution but also because small simulation boxes lack large-scale density fluctuations. As a consequence, not-large-enough simulation boxes would not allow the formation of large galaxy clusters, therefore hampering the growth of central (but also noncentral) galaxies. Thus, we mainly focus on the larger simulation boxes TNG100-1 and RefL0100N1504.

In the following, we compare the stellar mass buildup as predicted by the $\Lambda$CDM simulations with the masses of the observed high-redshift galaxy candidates ID~1514 \citep{Adams_2022}, ID~14924 \citep{Labbe_2022}\footnote{We only show ID~1514 and not the more massive candidate ID~14924 in the bottom right panel of Figure~\ref{figure_stellar_mass_distribution_LCDM} in order to be more conservative.}, GL-z11, GL-z13 \citep{Naidu_2022}, and CEERS-1749 \citep{Naidu_2022b_dust}. For the TNG runs, ID~1514 located at $z_{\mathrm{phot}} = 9.85_{-0.12}^{+0.18}$ and ID~14924 at $z_{\mathrm{phot}} = 9.92$ are compared with simulated galaxies at $z=10.00$. GL-z11 at $10.9_{-0.4}^{+0.5}$, GL-z13 at $13.1_{-0.7}^{+0.8}$, and CEERS-1749 at $16.0_{-0.6}^{+0.6}$ are compared with the simulations at $z = 10.98$, $11.98$, and $14.99$, respectively. The comparison with GL-z13 and CEERS-1749 is therewith more conservative because snapshots corresponding to lower redshifts than observed are addressed allowing the galaxies to grow in stars for a longer time span than in the observed cases.

In TNG100-1, the GSMF at $z = 14.99$ reaches a maximum value of $\log_{10}(M_{*}/M_{\odot}) = 7.32$ being therewith $\approx 191$ times lower than the stellar mass of CEERS-1749 with $\log_{10}(M_{*}/M_{\odot}) = 9.6_{-0.2}^{+0.2}$. At $z = 11.98$ and $z=10.98$, the maximum stellar masses of subhalos are $\log_{10}(M_{*}/M_{\odot}) = 8.07$ and $8.43$, respectively, which is less massive than GL-z13 with $\log_{10}(M_{*}/M_{\odot}) = 9.0_{-0.4}^{+0.3}$ and GL-z11 with $\log_{10}(M_{*}/M_{\odot}) = 9.4_{-0.3}^{+0.3}$. The maximum value of $\log_{10}(M_{*}/M_{\odot}) = 8.73$ at $z = 10.00$ is significantly lower than the stellar mass of ID~1514, which has $\log_{10}(M_{*}/M_{\odot}) = 9.8_{-0.2}^{+0.2}$. The discrepancy becomes even more significant for ID~14924, which has $\log_{10}(M_{*}/M_{\odot}) = 10.93$ at $z_{\mathrm{phot}} = 9.92$.

The above stellar masses refer to all stellar particles bound to the considered subhalo depending therewith on the subhalo-finding algorithm. The identification of subhalos can be disturbed, e.g. by merger events, which frequently occur especially at high redshifts. Therefore, we also assess the maximum stellar masses of halos, which accounts for both the fact that the subhalo finder can split a galaxy in clumps underestimating the total mass of the galaxy, and for the inclusion of observationally unresolved satellite galaxies. The maximum stellar masses of halos are $\log_{10}(M_{*}/M_{\odot}) = 7.32$ ($7.05$), $8.10$ ($7.88$), $8.43$ ($8.24$), and $8.78$ ($8.59$) at $z = 14.99$, $11.98$, $10.98$, and $10.00$, in the TNG100-1 (TNG50-1) run, respectively. Thus, using the most massive halo instead of the most massive subhalo in terms of its stellar mass does not significantly effect the results of the TNG runs.

In the EAGLE runs, ID~1514/ID~14924 and CEERS-1749 are compared with the GSMF at $z = 9.99$ and $15.13$, respectively. Unfortunately, the EAGLE database does not list snapshots that match the observed redshifts of GL-z11 and GLz-13. Thus, the EAGLE analysis only focuses on ID~1514, ID~14924, and CEERS-1749. 

The RefL0050N0752 and RefL0100N1504 snapshots contain galaxies reaching up to  $\log_{10}(M_{*}/M_{\odot}) \approx 7.70$ at $z=15.13$ and $\log_{10}(M_{*}/M_{\odot}) \approx 9.06$ at $z=9.99$, which is $\approx 79$ and $\approx 5.5$ times lower than the observed stellar mass of CEERS-1749 and ID~1514, respectively. The stellar mass of ID~14924 is 74 times higher than the most massive simulated galaxy at $z = 9.99$.

\begin{figure}
    \includegraphics[width=\columnwidth]{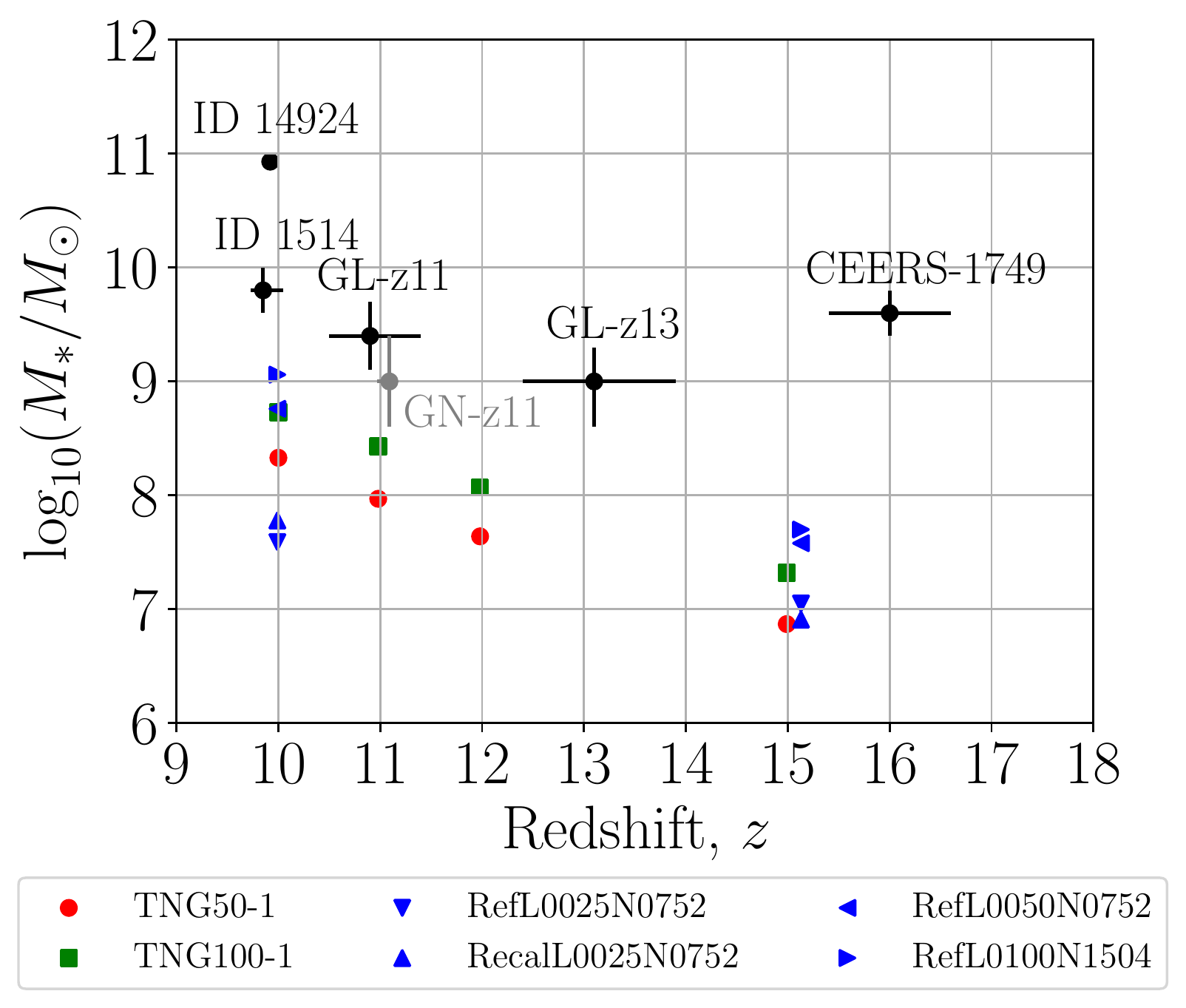}
    \caption{The most massive subhalo in terms of the stellar mass in dependence of redshift in the TNG50-1 (red), TNG100-1 (green), and EAGLE (blue) runs. The black error bars are the observed galaxy candidates by JWST as listed in Table~\ref{tab:comparison_simulation_observation}. The gray errobar shows GN-z11 \citep{Oesch_2016}.}
    \label{figure_Mstellar_redshift_simulation}
\end{figure}

The evolution of the stellar mass growth is summarized in Figure~\ref{figure_Mstellar_redshift_simulation} and Table~\ref{tab:comparison_simulation_observation} by showing the maximum stellar mass of a subhalo in dependence of redshift for different simulation runs. The observed high-redshift galaxy candidates are, by more than one order of magnitude, more massive than the most massive simulated galaxies in the $\Lambda$CDM framework.

The inferred stellar mass of observed galaxies is sensitive to the adopted star formation history (SFH) and initial mass function (IMF). In the following sections, we first investigate if the tension reported here of the stellar mass buildup in the early universe can be resolved if different SFHs of the observed galaxy candidates are assumed. Secondly, the effect of a varying IMF on the observed stellar masses is discussed.

\begin{deluxetable}{ccCrlc}[b!]
\tablecaption{Comparison of several observed galaxy candidates with $\Lambda$CDM simulations. \label{tab:comparison_simulation_observation}}
\tablecolumns{3}
\tablewidth{0pt}
\tablehead{
\colhead{ } &
\colhead{Redshift ($z$)} &
\colhead{$\log_{10}(M_{*}/M_{\odot})$} }
\startdata
CEERS-1749 & $16.0^{+0.6}_{-0.6}$ & $9.6^{+0.2}_{-0.2}$ \quad ($8.53$)\tablenotemark{a} \\
GL-z13 & $13.1^{+0.8}_{-0.7}$ & $9.0^{+0.3}_{-0.4}$ \quad ($8.01$)\tablenotemark{a} \\
GL-z11 & $10.9^{+0.5}_{-0.4}$ & $9.4^{+0.3}_{-0.3}$ \quad ($8.13$)\tablenotemark{a}\\
ID~1514 & $9.85^{+0.18}_{-0.12}$ & ${9.8}^{+0.2}_{-0.2}$ \quad ($8.53$)\tablenotemark{a}\\
 \hline
TNG50-1 & 14.99 & 6.87 \\
TNG50-1 & 11.98 & 7.64 \\
TNG50-1 & 10.98 & 7.97 \\
TNG50-1 & 10.00 & 8.33 \\
TNG100-1 & 14.99 & 7.32 \\
TNG100-1 & 11.98 & 8.07 \\
TNG100-1 & 10.98 & 8.43 \\
TNG100-1 & 10.00 & 8.73 \\
RefL0025N0752 & 15.13 & 7.05 \\
RefL0025N0752 & 9.99 & 7.59 \\
RecalL0025N0752 & 15.13 & 6.91 \\
RecalL0025N0752 & 9.99 & 7.78 \\
RefL0050N0752 & 15.13 & 7.58 \\
RefL0050N0752 & 9.99 & 8.76 \\
RefL0100N1504 & 15.13 & 7.70 \\
RefL0100N1504 & 9.99 & 9.06 \\
\enddata
\tablenotetext{a}{Lowest possible stellar mass value for an invariant canonical IMF as quantified in Section~\ref{subsec:minimum stellar mass for SFHs}}
\tablecomments{The first four rows show the photometric redshifts and stellar masses of the observed galaxy {red}{candidates} ID~1514 \citep[see table~3 and 5 of][]{Adams_2022}, GL-z11, GL-z13 \citep[see table~3 of][]{Naidu_2022}, and CEERS-1749 \citep[see table~3 of][]{Naidu_2022b_dust}. The other rows list the maximum stellar mass of a subhalo at a given redshift for the TNG and EAGLE runs. }
\end{deluxetable}

\subsection{
The minimum inferred galaxy masses for different star formation histories}
\label{subsec:minimum stellar mass for SFHs}

In order to calculate the minimum possible mass that the observed high-redshift galaxy candidates CEERS-1749, GL-z11, GL-z13, and ID~1514 can have for an invariant IMF, different sets of models of galaxies with different SFHs are constructed. 
Using stellar population synthesis models, we let the age of the modeled galaxies vary in the range of $\left[\approx 4, 400 \, \rm{Myr} \right ]$ to investigate their UV-band ($1500\,\rm{\AA}$) stellar mass (including remnants)-to-light ratio, $M_{*}/L_{\mathrm{UV}}$, for an invariant canonical IMF \citep[][]{Kroupa_2001,Kroupa_2013}. The lower limit is set by the implemented stellar evolution tracks of the Padova group \citep[][see also \citealt{Zonoozi_2019}]{Marigo_2007, Marigo_2008} and is roughly comparable to the mean stellar age ($\approx 1-20$~Myr) of observed high-redshift galaxies ($5 \la z_{\mathrm{spec}} \la 8$) as found by \citet{Carnall_2022_age}.

Since the spectral energy distribution fitting analysis of high-redshift star-forming galaxies shows more consistency with increasing SFHs, here we adopt the delayed-$\tau$ model \citep[e.g.,][]{Kroupa_2020}
\begin{equation}
\psi_{\mathrm{del}}(t)=\psi_{0} t \exp(-t/\tau) \, , 
\label{eq:delayed_tau_SFH}
\end{equation}
and an exponentially increasing SFH
\begin{equation}
\psi_{\mathrm{exp}}(t)=\psi_{0} \exp(t/\tau) \, ,
\label{eq:exponential_SFH}
\end{equation}
where $\psi(t)$ is the star formation rate (SFR), $t$ is the age since star formation started, $\psi_{0}$ is the normalization parameter, and $\tau$ is the e-folding time scale. The mass and light of a galaxy are calculated by an integral over the SFR. Note that, using the invariant IMF, at a given $t$, the mass-to-light ratio is independent of the total mass that is converted into stars. This is because of cancellation of the normalization parameter, $\psi_{0}$.
  
\begin{figure*}
\includegraphics[width=\linewidth]{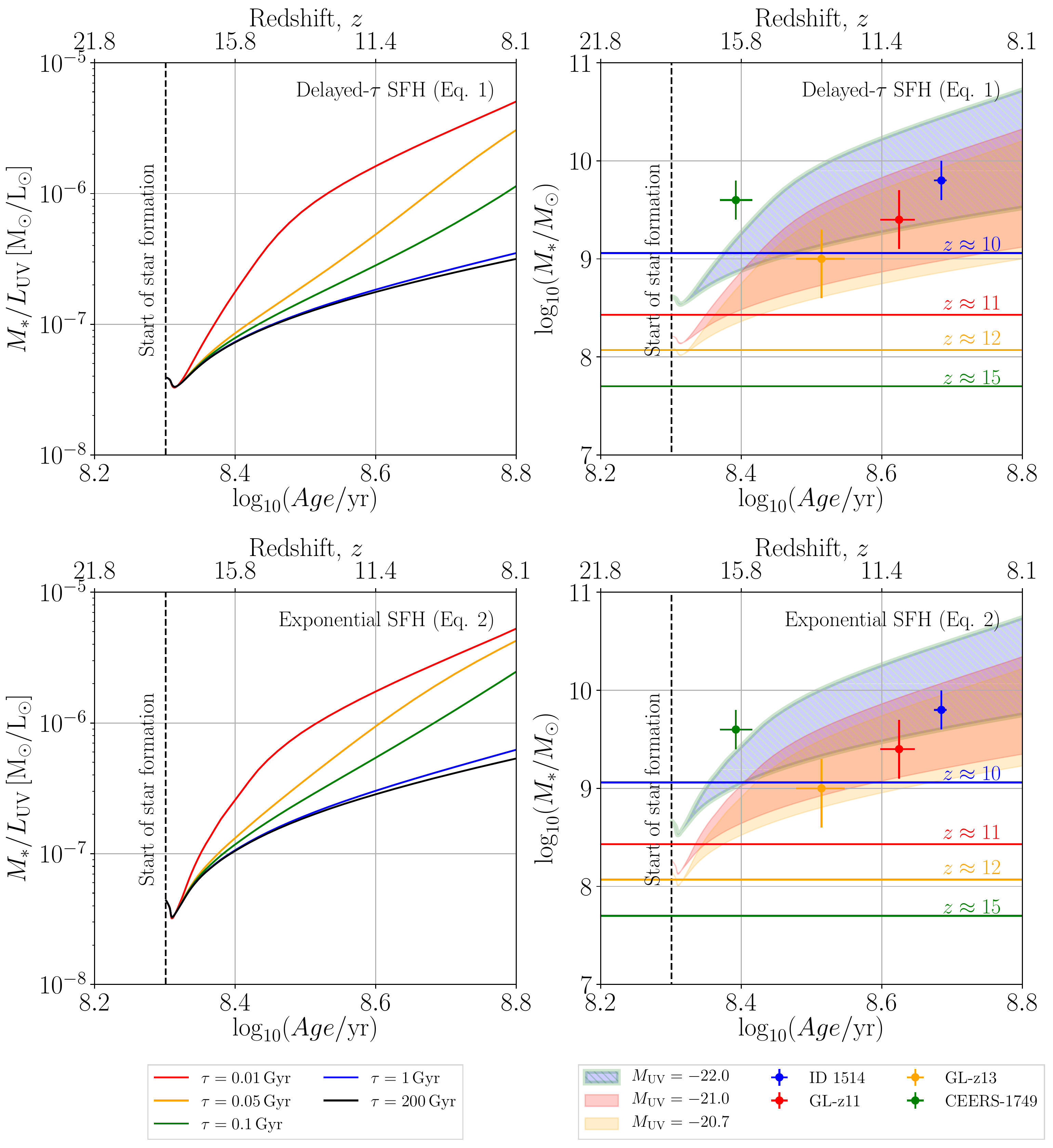}
\caption{Left panels: cosmic time evolution of the stellar $M_{*}/L_{\mathrm{UV}}$ ratio for stellar populations constructed assuming an invariant canonical IMF and using a delayed-$\tau $ (Eq.~\ref{eq:delayed_tau_SFH}; top panels) and an exponentially increasing SFH (Eq.~\ref{eq:exponential_SFH}; bottom panels). We assume that star formation starts $200$~Myr after the Big Bang (dashed vertical line). The minimum mass-to-light ratio for these galaxies assuming different SFHs is $M_{*}/L_{\mathrm{UV}}\approx 3.2 \times 10^{-8}\,M_{\odot}/L_{\odot}$. Right panels: cosmic time evolution of the total stellar mass of the galaxy candidates CEERS-1749 (green hatched area), GL-z13 (orange area), GL-z11 (red area), and ID 1514 (blue area) calculated based on the inferred mass-to-light ratios of the left panels. The colored areas cover the stellar mass range for SFHs with $\tau$ values between $0.01$~Gyr (upper limit) and $200$~Gyr (lower limit; see the left panels). The filled circles with error bars show the observed values as quantified by \citet{Adams_2022}, \citet{Naidu_2022b_dust}, and \citet{Naidu_2022}. The most massive subhalos in terms of stellar mass in the $\Lambda$CDM simulations are shown as horizontal lines.}
\label{Mass-time}
\end{figure*}

The effect of the SFHs on the $M_{*}/L_{\mathrm{UV}}$ ratio of galaxies by adopting different values of $\tau$ is shown in the left panels of Figure~\ref{Mass-time}.
We set that galaxies start forming stars 200~Myr after the Big Bang, with an averaged metallicity of [Fe/H] = $-2$, and we assume that the mass loss from galaxies is only through stellar evolution in the form of ejected gas. Since the stellar loss due to dynamical evolution is significant only for systems with initial stellar mass less than $M_{*}=10^6 M_{\odot}$, no stars are lost by the dynamical evolution of galaxies. As can be seen, the minimum mass-to-light ratio that can be considered for these galaxies assuming different SFHs is $M_{*}/L_{\mathrm{UV}}\approx3.2 \times 10^{-8}\,M_{\odot}/L_{\odot}$.   

According to the estimated UV absolute magnitude of these objects, $M_{\mathrm{UV,CEERS-1749}} = -22.0$, $M_{\mathrm{UV,GL-z13}}=-20.7$, $M_{\mathrm{UV,GL-z11}}=-21.0$, and $M_{\mathrm{UV,ID 1514}}=-22.0\,\rm{mag}$, we obtain the lowest possible stellar masses of $\log_{10}(M_{*}/M_{\odot})=8.53, 8.01, 8.13$, and $8.53$ for CEERS-1749, GL-z13, GL-z11, and ID~1514 in the case of an exponential SFH, respectively, as visualized in the right panels of Figure~\ref{Mass-time}. 

These lower stellar mass limits just resolve the discrepancy for GL-z11, GL-z13, and ID 1514 (see the vertical dashed-dotted lines in Figure~\ref{figure_stellar_mass_distribution_LCDM}). In the case of CEERS-1749, the maximum stellar mass obtained in the $\Lambda$CDM simulation is $\approx6.8$ lower than its inferred lower limit. 

\subsection{Galaxy masses for a varying IMF}
\label{subsec:stellar mass for IGIMF}
In the previous section we applied an invariant IMF but recent observations \citep[e.g.,][]{Schneider_2018,Zhang_2018,Senchyna_2021} suggest that the mass distribution of a stellar population may depend on its local star forming environment. Especially metal-poor Population III stars are expected to follow a top-heavy IMF. 

A theoretical framework to describe the stellar population of an entire galaxy is the integrated galactic initial mass function (IGIMF) theory, which adds up all the IMFs of starforming regions (embedded clusters) within a galaxy \citep[][]{Kroupa_2003,Weidner_2006}. The resulting galaxy-wide IMF (gwIMF) systematically varies with the global SFR and averaged metallicity of the galaxy, i.e. the gwIMF becomes top-heavy for galaxies with $SFR \ga 1 \, M_{\odot}\,yr^{-1}$ and metallicities [Fe/H] $< 0$ \citep[see figure~2 of][]{Jerabkova_2018_IGIMF} compared to the canonical IMF. 

\begin{figure*}
\includegraphics[width=\linewidth]{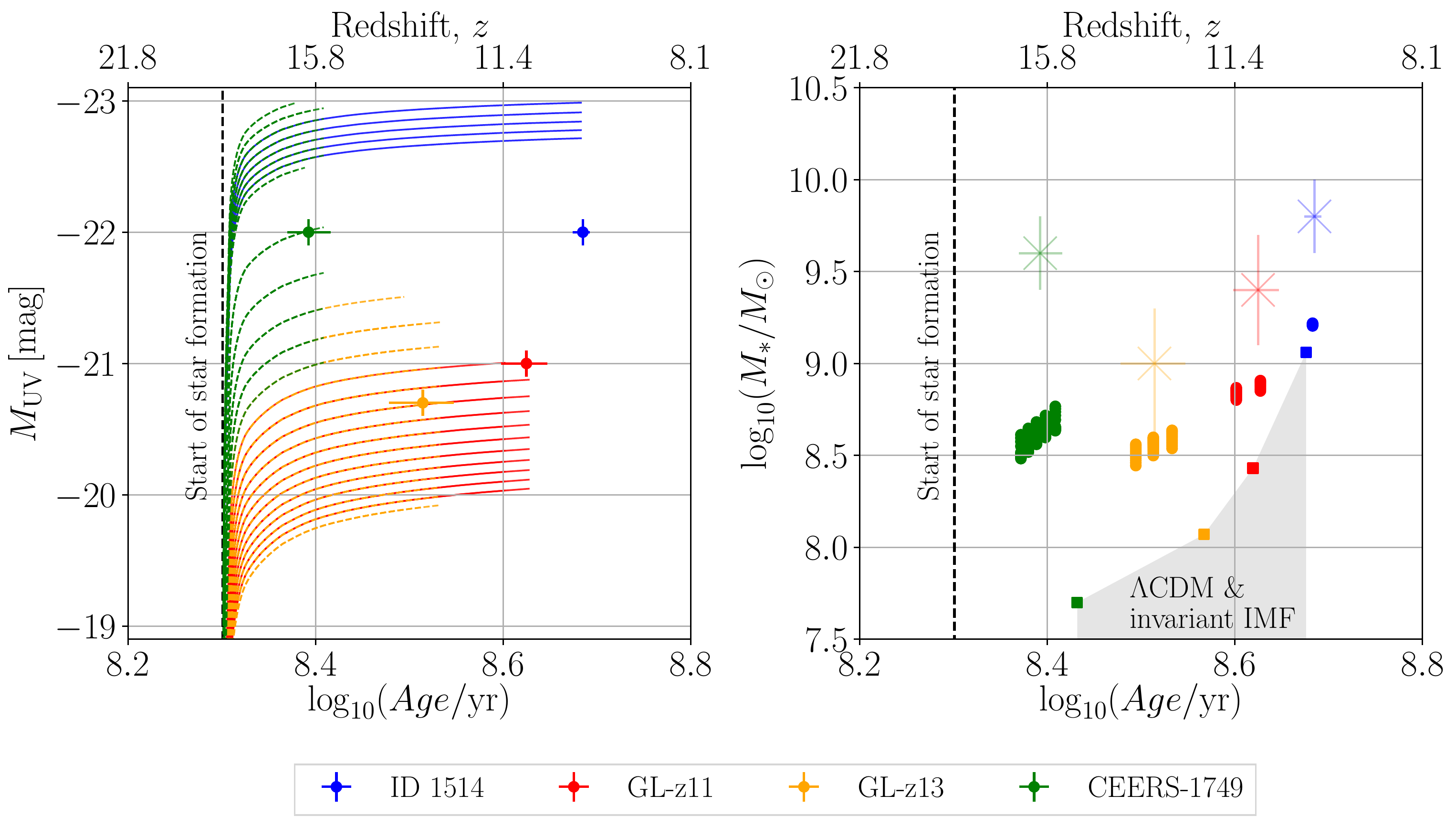}
\caption{Left panel: cosmic time evolution of the absolute UV-band magnitude for stellar populations constructed using constant SFRs and assuming the IGIMF theory and that star formation starts $200$~Myr after the Big Bang (dashed vertical line). The shown green, orange, red, and blue models match the observed magnitudes within $\pm 1$~mag at the $1 \sigma$ redshift interval of the galaxy candidates CEERS-1749, GL-z13, GL-z11, and ID~1514, respectively. The error bars show the observed absolute magnitudes of these four galaxy candidates. Right panel: the filled circles show the stellar masses of the galaxy candidates (of same color as the corresponding filled circles) assuming the $M_{*}/L_{\mathrm{UV}}$ ratios of the IGIMF models presented in the left panel. The faded error bars refer to the reported stellar masses as derived by \citet{Adams_2022}, \citet{Naidu_2022}, and \citet{Naidu_2022b_dust} based on an invariant canonical IMF. The squares mark the upper stellar mass limit found in the $\Lambda$CDM simulations assuming an invariant canonical IMF, and would be smaller for stronger-feedback regulation through an early top-heavy gwIMF.}
\label{figure_IGIMF}
\end{figure*}

In order to study the effect of a varying IMF on the high-redshift galaxy candidates, we calculate the stellar population of galaxies assuming the latest IGIMF formalism by \citet{Yan_2021} and using an IGIMF Fortran code developed by Akram Hasani Zonoozi. For simplification, we assume that all galaxies start to form stars $200$~Myr after the Big Bang, with a constant SFR over time, and an average metallicity of [Fe/H] = $-2$. Since the gwIMF is time dependent, we require that realistic IGIMF models have to match the observed $M_{\mathrm{UV}}$ within $\pm 1$~mag in the $1 \sigma$ interval of the observed redshift of the corresponding galaxy candidate. The left panel of Figure~\ref{figure_IGIMF} shows the time evolution of the absolute UV-band magnitude for IGIMF models that fulfill these constraints for observed galaxy candidates CEERS-1749, GL-z13, GL-z11, and ID~1514. These models have constant SFRs in the range of $\approx 2 - 30 \,M_{\odot}\,yr^{-1}$ and thus a top-heavy IMF. The stellar masses at the observed redshifts of the galaxy candidates are shown in the right panel of Figure~\ref{figure_IGIMF} and are systematically lower than the derived stellar masses of \citet{Adams_2022}, \citet{Naidu_2022}, and \citet{Naidu_2022b_dust} because of the top-heavy gwIMF compared to the canonical IMF.

\section{Discussion and Conclusion} \label{sec:Discussion and Conclusion}
While the redshifts of the galaxy candidates need to be spectroscopically verified, we use these JWST observations to quantify how quickly galaxies form in the currently most advanced cosmological simulations. Using state-of-the-art $\Lambda$CDM simulations of the IllustrisTNG and EAGLE project, we showed that the stellar mass buildup is much more efficient in the early universe than predicted by these $\Lambda$CDM models \citep[see also, e.g.,][]{BoylanKolchin_2022, Lovell_2022}. In particular, the stellar masses of ID~1514 \citep{Adams_2022}, ID~14924 \citep{Labbe_2022}, GL-z11, GL-z13 \citep{Naidu_2022}, and CEERS-1749 \citep{Naidu_2022b_dust} analyzed in Section~\ref{sec:Results} are higher by about one order of magnitude than the most massive galaxies formed in these simulations. In particular, massive high-redshift candidates appear more frequent at $z \ga 10$ than expected in the $\Lambda$CDM framework. For example, \citet{BoylanKolchin_2022} argued that a volume of $\approx 10^{8}\,\rm{cMpc^{3}}$ is required to explain ID~14924 with $\log_{10}(M/M_{\odot}) = 10.93$ at $z = 9.92$ \citep{Labbe_2022}. However, the survey covers $\approx 10^{5}\,\rm{cMpc^{3}}$ at $z = 10 \pm 1$ \citep[see section~3 of][]{BoylanKolchin_2022}. The TNG100-1 and RefL0100N1504 simulations have a box volume of $\approx 10^6\,\rm{cMpc^{3}}$ suggesting that the absence of massive galaxies in these runs is not because of a too small simulation volume.

The discrepancy between the observed and simulated stellar mass buildup could be caused by several reasons. First of all, high photometric redshifts can emerge due to dust reddening. For example, \citet{Zavala_2022} demonstrated that Lyman-break galaxy candidates at $z_{\mathrm{phot}} \ga 12$ can resemble dusty star-forming galaxies at $z \la 6-7$. Secondly, it could be that the high observed stellar masses are caused by an erroneous calibration of JWST.

Furthermore, it has been argued that star formation could be much more efficient in the early universe \citep[e.g.,][]{Naidu_2022,Harikane_2022,Mason_2022}. Assuming that the $\Lambda$CDM model is the correct description of the Universe, this would mean that the underlying galaxy formation and evolution ISM models of the EAGLE and IllustrisTNG runs must be improved in order to reproduce such galaxies. For example, these simulations assume that gas above a given density threshold \citep[e.g.,][]{Schaye_2015,Nelson_2019} is able to form stars, which is likely a too simplified implementation especially for describing high-redshift galaxies. \citet{BoylanKolchin_2022} showed that even a $100\%$ star formation efficiency in $\Lambda$CDM would not be enough to explain the stellar mass density measured by \citet{Labbe_2022}.

Another possibility is that the IMF systematically varies with the galactic properties. The IllustrisTNG and EAGLE simulations and the analysis of the Sections~\ref{sec:Results} and \ref{subsec:minimum stellar mass for SFHs} assume an invariant IMF, but it is expected that metal-poor starforming stellar populations follow a top-heavy IMF. Using the IGIMF theory we calculated the gwIMF of the observed galaxy candidates in dependence of the metallicity and SFR of forming galaxies resulting in lower stellar masses compared to an invariant canonical IMF. For this, the IGIMF theory has to be included in cosmological simulations \citep[see, e.g.,][]{Ploeckinger_2014} in order to make a firm conclusion if a top-heavy IMF can resolve the reported tension.

Finally, the present findings can also imply that structure formation is much more efficient and/or that the observed universe is even older than predicted by $\Lambda$CDM. The existence of these massive galaxies $\approx 300-400$~Myr after the Big Bang also questions the hierarchical (bottom-up) structure formation suggesting that late-type galaxies begin to form early through the initial monolithic collapse of rotating post-Big-Bang gas clouds \citep{Wittenburg_2020} while early-type massive galaxies and associated formation of supermassive black halos form by the monolithic collapse of post-Big-Bang gas clouds with little net rotation \citep[e.g.,][]{Kroupa_2020_SMBH,Wittenburg_2020,Yan_2021,Eappen_2022}.

Evidence for an enhanced growth of structures has been reported at different astrophysical scales and redshift ranges in the observed Universe. For example, \citet{Steinhardt_2016} showed that the observed number density of luminous galaxies at $5 \la z \la 10$ is much higher than predicted by the $\Lambda$CDM model (see their figure~1). However, their analysis relies on the stellar-to-halo mass relation from \citet{Leauthaud_2012} measured only at $z=0.2-1$, while, e.g., \citet{Behroozi_2019} suggest that there is a strong evolution at $z\ga5$. Furthermore, the existence of the massive interacting galaxy cluster El Gordo \citep[ACT-CL J0102-4915;][]{Marriage_2011} at $z = 0.87$ and the Keenan-Barger-Cowie void \citep[][]{Keenan_2013} at $z \la 0.07$ both individually falsify the hierarchical $\Lambda$CDM structure formation with more than $5 \sigma$ \citep{Haslbauer_2020,Asencio_2021}.

An enhanced growth of structure compared to the $\Lambda$CDM paradigm is expected in Milgromian dynamics \citep{Milgrom_1983,Angus_2009,Malekjani_2009,Famaey_2012,Kroupa_2012,Haslbauer_2020,Banik_2022_MONDreview}. Assuming the cosmic microwave background as the $z = 1100$ boundary condition and because of the reduced power on $< 1\,\rm{Mpc}$ scales compared to $\Lambda$CDM \citep{Angus_2011} due to the missing cold dark matter (CDM) component, it may be impossible to form galaxies in the early universe.

This work indicates that the currently available most advanced $\Lambda$CDM simulations cannot form galaxies as massive as observed at $z_{\mathrm{phot}} \ga 10$. This tension needs to be readdressed for extreme SFHs and/or if the gwIMF was top-heavy which would reduce the stellar mass buildup through more intense feedback.

Upcoming ultradeep and wider-area JWST observations will reveal more light on the number density of such luminous high-redshift galaxies over redshift required to evaluate the significance of the here-reported tension of the stellar mass buildup of high-redshift galaxies in more detail. 

\acknowledgments
We thank an anonymous referee for helpful comments that significantly improved the manuscript. This project was largely conducted at the Charles University in Prague and we acknowledge the ``DAAD-Eastern-European" exchange program for financing visits at the Charles University in Prague during which we discussed several aspects of the formation and evolution of high-redshift galaxy candidates in the $\Lambda$CDM and Milgromian frameworks. We are also grateful to Nils Wittenburg, Nick Samaras, Ingo Thies, and Elena Asencio for helpful discussions on structure formation in Milgromian dynamics.

\bibliographystyle{aasjournal}
\bibliography{JWST_LCDM}

\end{document}